\documentclass[%
 aip,
 amsmath,amssymb,
 reprint, 
twocolumn 
]{revtex4-1}

\usepackage{graphicx}% Include figure files
\usepackage{dcolumn}% Align table columns on decimal point
\usepackage{bm}% bold math
\usepackage{multirow}
\usepackage[utf8]{inputenc}
\usepackage[T1]{fontenc}
\usepackage{mathptmx}
\usepackage{etoolbox}
\usepackage{subcaption}
\usepackage{graphicx}
\usepackage[dvipsnames]{xcolor}
\newcommand{\ym}[1]{#1}
\newcommand{\ny}[1]{#1}
\newcommand{\nycm}[1]{}
\newcommand{\nynd}[1]{#1}
\newcommand{\nyrd}[1]{#1}
\newcommand{\nycmnd}[1]{}
\newcommand{\newversion}[1]{#1}

\makeatother
\begin{document}

\preprint{AIP/123-QED}

\title[Jinzhen Zhu]{A unified framework for coarse grained molecular dynamics of proteins with high-fidelity reconstruction}
% Force line breaks with \\
\author{Jinzhen Zhu}

\email{zhujinzhenlmu@gmail.com,zhujinzhen@pjlab.org.cn}
\affiliation{ 
Shanghai AI Laboratory, Shanghai, 200030, China
}
\date{\today}% It is always \today, today,
\begin{abstract}
Simulating large proteins using traditional molecular dynamics (MD) is computationally demanding. 
To address this challenge, we propose a novel tree-structured coarse-grained model that efficiently captures protein dynamics.
By leveraging a hierarchical protein representation, our model accurately reconstructs high-resolution protein structures, with sub-angstrom precision achieved for a 168-amino acid protein.
We combine this coarse-grained model with a deep learning framework based on stochastic differential equations (SDEs).
A neural network is trained to model the drift force, while a RealNVP-based noise generator approximates the stochastic component.
This approach enables a significant speedup of over 20,000 times compared to traditional MD, allowing for the generation of microsecond-long trajectories within a few minutes and providing valuable insights into protein behavior.
Our method demonstrates high accuracy, achieving sub-angstrom reconstruction for short (25 ns) trajectories and maintaining statistical consistency across multiple independent simulations.
\end{abstract}
\maketitle

\section{\label{sec:introduction}Introduction}
Molecular dynamics simulations are a vital tool for protein structure prediction and drug design\cite{Hollingsworth2018}. However, simulating the motions of large biological molecules with traditional MD methods is computationally expensive, requiring significant time and resources due to factors like system size and simulation timescale\cite{Zimmerman2016}. Accelerating these computations is a key focus for computational chemists and biologists. Efforts are primarily directed towards two approaches: hardware and software improvements.
In the hardware arena, traditional MD codes have been rewritten to leverage the parallel processing power of Graphics Processing Units (GPUs)\cite{Bergdorf2021}. Additionally, the development of supercomputers like Anton has significantly boosted computational speed\cite{Shaw2021}.
On the software front, researchers employ enhanced sampling methods, such as replica exchange methods, to explore a wider range of configurations within the simulated system\cite{Bernardi2015}. Additionally, coarse-grained MD simulations focus on capturing essential features of the system while averaging over less crucial details. This approach offers significant computational and conceptual advantages compared to more detailed all-atom models.
\par
Coarse-grained methods face several challenges due to the inherent complexity of biological molecules like proteins. These challenges include capturing the intricate motions of protein backbones due to the presence of flexible side chains, loops, and disulfide bonds. Additionally, describing an implicit solvent environment that accurately reflects the crucial role of water molecules in stabilizing folded protein structures, particularly hydrophobic interactions, proves difficult with CG methods.
To overcome these challenges, CG methods employ two main paradigms: "top-down" and "bottom-up".\cite{Noid2013}.
The "top-down" CG method constructs collective variables (CVs) based on experimental observations, often leveraging thermodynamic data.
In contrast, the bottom-up approach utilizes classical or empirical atomic models. It requires a deeper understanding of the underlying physics and chemistry. Here, researchers leverage knowledge of fundamental properties like bond lengths, bond angles, dihedral angles, and non-bonded interactions to derive a coarse-grained representation of the system. 
\par
Within the bottom-up paradigm of CG methods, researchers leverage the framework of statistical mechanics to define effective interactions, such as potential energies\cite{Liwo2001}.
Over the past decades, several traditional coarse-grained methods have emerged, demonstrating significant improvements in capturing protein dynamics\cite{Mim2012,Chu2005,Yu2021,Toth_2007,Li2010}.
The incorporation of machine learning (ML) techniques has further enhanced the performance of CG simulations\cite{Yu2021,Wang2019,Husic2020}.
For instance, Zhang et al. employed deep neural networks (DNNs) to approximate potential energy surfaces from coarse-grained coordinates, achieving consistent distributions of CVs compared to those obtained from original MD simulations\cite{Zhang2018}.
This framework further enabled enhanced sampling of large proteins \cite{Wang2022}.
\par
The three-dimensional structure of a protein, crucial for its function, is determined by a interplay between fixed covalent bonds and flexible internal angles. Among these internal angles, torsion angles (also known as dihedral angles)  exhibit a more significant influence on protein conformation compared to the relatively static bond angles. Torsion angles display substantial fluctuations during MD simulations, significantly impacting how a protein folds. Consequently, these torsion angles are frequently incorporated into CG representations.
Electron orbital hybridization dictates the preferred geometries of bond angles (e.g., $sp^3$ or $sp^2$). 
However, experimental observations and MD simulations reveal slight deviations from these ideal values. 
These deviations can be particularly significant when attempting to reconstruct a protein structure solely based on torsion angles. 
The error accumulates with each amino acid residue, as the position of an amino acid is dictated by the conformation of the preceding residues in the polypeptide chain.
Therefore, CG representations that solely employ torsion angles are useful for obtaining a general understanding of a protein's backbone conformation or predicting side-chain structures (as demonstrated in references\cite{Zhang2018,Xu2022,Xu2023}.
However, they are inadequate for high-precision tasks such as complete protein structure prediction. For instance, the AlphaFold 2 (AF2) framework utilizes torsion angles for side-chain prediction but employs Cartesian coordinates for backbone prediction\cite{Jumper2021}.
Recent work by Kohler et al. highlights the potential benefits of incorporating bond angles. Their CG model utilizes both torsion and bond angles for the central carbon (CA) atoms, successfully capturing protein folding/unfolding transitions in small systems\cite{Kohler2023}.
However, this approach does not capture the detailed dynamics of all the atoms and side chains within a protein.
\par
In this research, we present a framework for constructing a bidirectional map between protein Cartesian coordinates and a minimal set of interpretable coarse-grained CVs. This framework incorporates all heavy atoms in the backbone and side chains, enabling a comprehensive representation of the protein structure. Notably, our approach allows MD simulations to be performed solely using the CVs, eliminating the need for explicit Cartesian coordinates. This significantly accelerates simulations while maintaining consistent performance with all-atom MD simulations, as measured by CV distributions and RMSDs at most time steps. Compared to existing methods that achieve only overall statistical consistency on CVs\cite{Zhang2018}, our framework offers the advantage extra high accuracy in time series of RMSDs. 
\section{\label{sec:methods}Methods}
\subsection{\label{sec:methods-cv}Collective variables}
% \begin{equation}
%     \hat{R}}(\hat{u},\theta)=\hat{I}}\cos\theta+\hat{u}\hat{u}^T(1-\cos\theta)+\tilde{u}\sin\theta
% \end{equation}
\subsubsection{\label{sec:level3}coordinate transformation matrix}
\par
Our protein representation begins with a general description of atomic coordinate hierarchy in chemical compounds.
For clarity, this work adopts a non-standard notation. 
The symbol $\psi$ will represent any dihedral angle and $\varphi$ will represent any bond angles.
According to \ny{Parsons et al.}\cite{Parsons2005}, the operating matrix $\hat{R}(\hat{u},\theta)$ for rotating $\theta$ along a normalized rotation vector $\vec{u}=[\vec{u}_x,\vec{u}_y,\vec{u}_z]^T$ is
\begin{equation}
    \begin{bmatrix}
\vec{u}_x^2(1-\text{c}\theta)+\text{c}\theta & \vec{u}_x\vec{u}_y(1-\text{c}\theta)-\vec{u}_z\text{s}\theta  & \vec{u}_x\vec{u}_z(1-\text{c}\theta)+\vec{u}_y\text{s}\theta\\ 
\vec{u}_x\vec{u}_y(1-\text{c}\theta)+\vec{u}_z\text{s}\theta & \vec{u}_y^2(1-\text{c}\theta)+\text{c}\theta  & \vec{u}_y\vec{u}_z(1-\text{c}\theta)-\vec{u}_x\text{s}\theta\\ 
\vec{u}_x\vec{u}_z(1-\text{c}\theta)-\vec{u}_y\text{s}\theta & \vec{u}_y\vec{u}_z(1-\text{c}\theta)+\vec{u}_x\text{s}\theta  & 
\vec{u}_z^2(1-\text{c}\theta)+\text{c}\theta\end{bmatrix}
\end{equation}
where \ny{$\text{s}\theta=\sin\theta$ and $\text{c}\theta=\cos\theta$}.
Considering the transformation, the general formula of coordinate transformation goes as a $4\times4$ matrix, where $\vec{\bold{0}}$ is a \ny{length-3 vector} filled with zeros and $\vec{T}$ is a \ny{length-3 vector} representing the translation:
\begin{equation}
\hat{M}(\vec{u},\theta,\vec{T})=\begin{bmatrix}
\hat{R}(\hat{u},\theta) & \vec{T} \\ 
\vec{\bold{0}}^T & 1
\end{bmatrix}.
\end{equation}
\par
\begin{figure}
\centering
\begin{subfigure}{0.48\textwidth}
    \includegraphics[width=\textwidth]{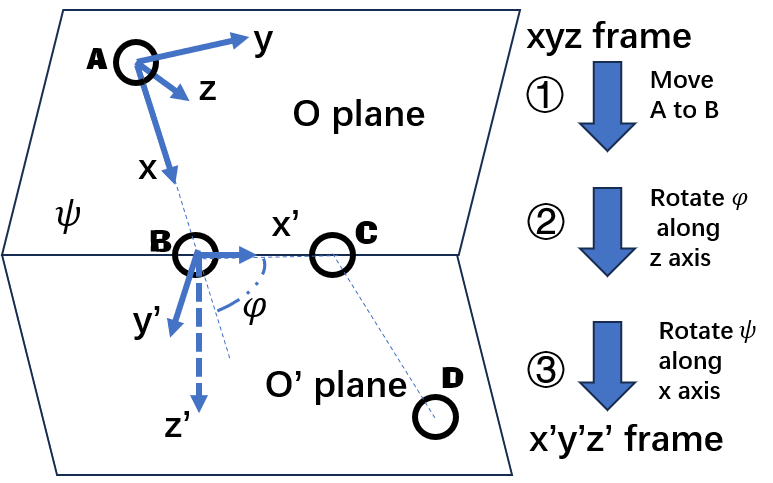}
    % \caption{Firts subfigure.}
\end{subfigure}
\begin{subfigure}{0.48\textwidth}
    \includegraphics[width=\textwidth]{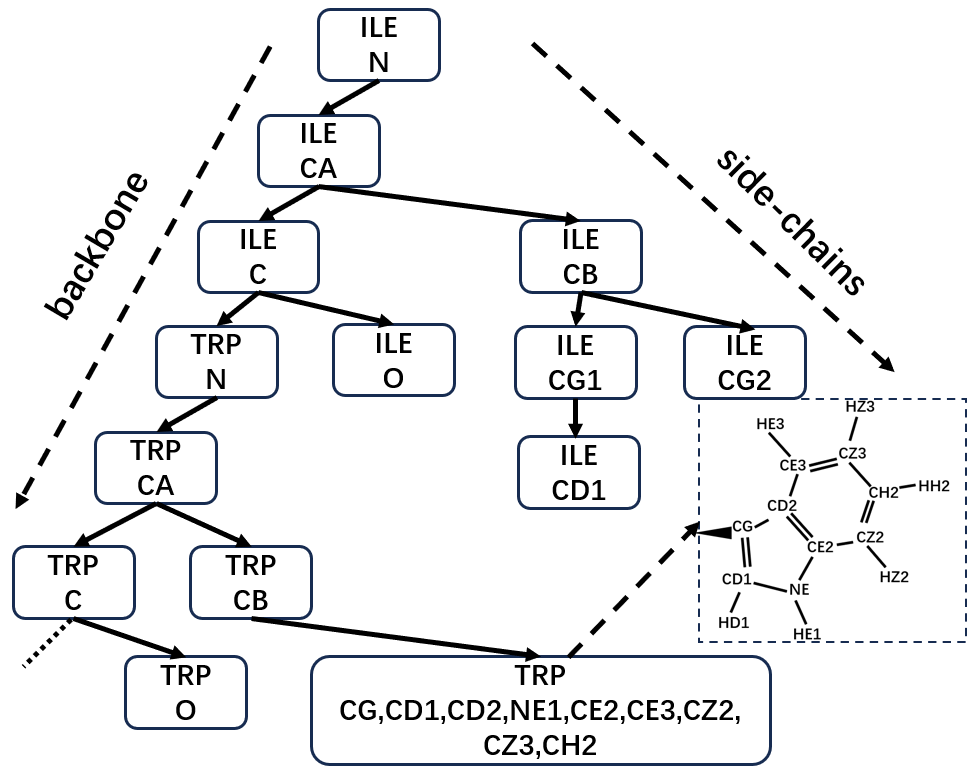}
    % \caption{Firts subfigure.}
\end{subfigure}
\caption{This illustration depicts coordinate transform (upper) and tree structure of CVs of atoms in proteins (lower) illustration.
Top panel presents four atoms (A, B, C, and D) that define two planes: O (formed by A, B, and C) and O' (formed by B, C, and D). The two planes are described by xy plane of global coordinate system $xyz$ and local coordinate system $x'y'z'$, respectively. The dihedral angle $\psi$, and the bond angle $\pi-\varphi$ describe coordinate transformation \nynd{as shown in Eq.~\ref{eq:hatO}. In the lower panel, an artificial protein with first two amino acids being ILE and TRP is applied to illustrate the tree-structure.} 
Each node stores atoms of an amino acid.
The structure of amino acids could be found in Fig.~\ref{fig:ile-trp} in the appendix.}
\nycmnd{Some nodes have only one atom and some have more than one, see the node at the right left bottom.}
\label{fig:illustration}
\end{figure}

Converting the coordinates \ny{from the} local axes frame (x'y'z') to parent axes (xyz) can be understood as transforming the parent axes data (axes vector and the origin) to the the local axes data in the parent axes frame, as illustrated in Fig.~\ref{fig:illustration}.
This process includes three steps with operator written as 
\begin{equation}\label{eq:hatO}
    \hat{O}=\hat{M}_3\hat{M}_2\hat{M}_1\ny{,}
\end{equation}
where
first, $\hat{M}_1$ translates the frame along x axis by bond length $l_b$ \ny{with} $\hat{R}(\hat{u},\theta)=\hat{I}$ \nynd{with $\hat{I}$ being the identity matrix},  and $\vec{T}=[0,0,l_b]^T$;
second, $\hat{M}_2$ rotates the frame along the new z axis by bond angle with $\theta=\varphi$, $\vec{T}=\vec{0}$ and $\vec{u}=[0,0,1]^T$;
third, $\hat{M}_3$ rotates the frame along the new x axis by dihedral angle with $\theta=\psi$, $\vec{T}=\vec{0}$ and $\vec{u}=[1,0,0]^T$. 
\ny{These three steps are also illustrated as \textcircled{1}, \textcircled{2} and \textcircled{3}  in the upper pannel of Fig.~\ref{fig:illustration}}
\par
For complex molecules like proteins, multiple coordinate frames are used. One frame serves as the "global" reference, and data from other frames are transformed to this reference frame.
The operator for obtaining the coordinates in current local frame to the given frame is denoted as 
\begin{equation}
    \hat{P}_{I,J}=\Pi_{i=I+1}^{J}\hat{O}_i,1\leq I<J
\end{equation}
with index showing the coordinate frames before current frame, and $I$ and $J$ being the \ny{parent and local} frame index.
\ny{$\hat{O}_i$ is $\hat{O}$ from Eq.~\ref{eq:hatO} in the $i$th coordinate frame.}
We denote the global operator for converting current coordinates to the "global" reference as
\nynd{
\begin{equation}\label{eq:ppmin1}
    \hat{P}_I\equiv\hat{P}_{I,1}=\hat{P}_{I-1}\hat{O}_I.
\end{equation}
}
This approach leverages a recursive formula to efficiently compute the global transformation matrix for each reference frame. This eliminates the need for repeated calculations on individual sub-coordinates, simplifying the computational process.
Consequently, equation 
\begin{equation}\label{eq:single-tree}
    \vec{X}=\hat{P}_I\vec{X^0},
    \vec{X}=[\vec{x}^T,1]^T,\vec{X^0}=[\vec{x^0}^T,1]^T
\end{equation}
can be readily applied to determine any protein atom's Cartesian coordinates within the global reference frame\nycm{the position of an atom in "global frame" or any other "local frames" are all represented in Cartesian coordinates during the transformation from CVs to final protein structure, in short, they are all Cartesian coordinates. But the values of these coordinates are different, depending on which frame you are using. In figure 1, B is (0,0,0) in local x'y'z' frame but is (bond,0,0) is global frame xyz (if we take xyz as the global frame).},
where vector $\vec{x}$ and $\vec{X}$ are original Cartesian coordinates ($3\times1$) and extended coordinates ($4\times1$).
For convenience, subscript $I$ indicates the number and index of coordinate transformation from the "global" coordinate.
A superscript of "0" denotes a "local" coordinate within its frame, otherwise it's a "global" coordinate.
As $\vec{x^0}$ are only determined by the amino type, the Cartesian coordinates $\vec{x}$ are only dependent on $\hat{P}_N$.
The bond lengths, $l_b$, used in the calculations can also be retrieved from the standard amino acid structures.
It is noteworthy that the \nynd{the protein backbone C-N bonds} have a fixed length of about 1.32 angstroms.
\par
The conversion between a protein's 3D structure (Cartesian coordinates) and its internal torsion and bond angles is straightforward and is neglected here. This establishes a bidirectional mapping, facilitating analysis from both perspectives.
\nynd{Our calculation only consider the heavy atoms for the sake of computational efficiency.}
Additionally, rigid moieties like benzene rings require only two angles for complete description. This significantly reduces the number of variables needed, streamlining computational demands.
While introducing some additional parameters compared to traditional methods, our approach offers substantial advantages. It allows reconstruction of the complete 3D structure, including side chains, leading to more accurate predictions of interactions – crucial for molecular simulations.

\subsubsection{\label{sec:methods-data-structure}Tree data structure}
\par
Recursive data structures naturally lend themselves to representation using tree structures. This approach has been successfully implemented in J. Zhu's prior work on quantum physics \cite{Zhu2021,Zhu2020,Zhu2020b}.
Similarly, the coordinates of each reference frame within a protein can be effectively represented as a tree structure, as demonstrated in Fig.~\ref{fig:illustration} using an artificial protein with first two amino acids being ILE and TRP as an example.
The two-amino acid peptides do not exist, but these artificial peptides are always used for illustration or testing purpose. To avoid any potential confusion, I change it to a peptide with first two amino acids being ILE-TRP.
This tree structure efficiently represents protein hierarchy and geometry. Each node serves as a local reference frame, storing the local coordinates of its constituent atoms.
Nodes inherit the transformation operator, 
denoted as $\hat{P}_I$ (refer to Eq.~\ref{eq:ppmin1}), from their parent nodes ($\hat{P}_{I-1}$). This inheritance reflects the hierarchical relationships within the protein.
Furthermore, the tree captures geometric information. For instance, atoms in TRP forming rigid rings (e.g., CG$\cdots$ CH2) reside within a single node.
For a more detailed breakdown of the hierarchy within each amino acid, please refer to Table~\ref{tab:amino-hierarchy} in the appendix.
This approach acknowledges their inherent rigidity and reduces the number of redundant parameters needed to define their relative positions, promoting both data storage efficiency and a more natural representation of the protein's structure.
\par
While the recursive multiplications of matrices can be computationally expensive, the reconstruction of protein 3D structures is only required during trajectory analysis. 
This means these computationally intensive operations are not necessary for real-time molecular dynamics simulations, thereby minimizing their impact on overall performance.
Furthermore, the reconstruction of CV trajectories can be efficiently parallelized across CPUs and GPUs. This distributed processing approach effectively mitigates the performance drawbacks associated with recursive computation time.
\subsection{\label{sec:method-ff}Propagation and molecular dynamics}
Simulating large biomolecular systems necessitates capturing their statistical behavior. \ny{There exist two main approaches: stochastic methods\cite{Harrison2009,GBhanot1988} and equilibrium sampling\cite{Karplus1990,Bussi2007}.
Equilibrium sampling uses an initial state that already reflects the desired statistical properties, followed by noise-free simulations\cite{Karplus1990}.
Stochastic methods introduce random noise at each simulation step to mimic statistical fluctuations, starting from a defined initial state. Examples include random walk\cite{Codling2008} and could be decribed by stochastic differential equations (SDE)\cite{Slavik2013}.
The SDE is also widely applied in computer science for generation of figures~\cite{Song2019,Song2021}, and later applied molecular dynamics schemes to generate MD trajectories\cite{Wu2023,Petersen2021}.
}
\par
This section explores the potential of leveraging strategically chosen data points with minimal correlation from MD trajectories to  partially mimic SDEs for statistical analysis\cite{Codling2008}. By constructing a dataset that reflects the system's exploration of phase space, we can gain relative information about its dynamics through the lens of an SDE. However, careful selection of the time interval between saved points is crucial, as large gaps can miss important events and hinder accurate analysis\cite{Slavik2013}. This approach offers a potentially faster and more efficient way to gain statistical insights into complex systems, particularly when traditional methods become computationally expensive.
Our work uses the strategy by utilizing a machine learning propagator trained on saved MD trajectories and this propagator directly predicts the \ny{coordinates of next snapshot}.
We will show that this propagator effectively includes the drift force and the noise part in the SDE.
\par
Our work builds upon recent advances in SDE-based machine learning, where models learn to reverse a data corruption process (diffusion) \cite{Ho2020,Nichol2021}. However, we propose a novel training scheme due to key differences in our setting.
First, data characteristics: Unlike prior works that employ artificially generated, noise-driven trajectories\cite{Ho2020,Nichol2021}, our model deals with physical, sequential data with inherent physical constraints.
Second, Process direction: We learn a forward process directly, contrasting with the reversed process learned in diffusion models.
Third, data scarcity: We operate with limited real-world data, as opposed to the large, artificially generated datasets used in previous works.
Computing the noises with limited trajectory data may include large numerical errors.
\subsubsection{SDE in collective variables}\label{sec:SDE-CV}
For all Cartesian coordinates transformation, Eq.~\ref{eq:single-tree}  can be written as 
\begin{equation}
    \mathit{R} =\mathcal{P}(\vec{\theta}, \mathit{R}^0)
\end{equation}
for N Cartesian coordinates
    $\mathit{R}=[\vec{x}_1,\vec{x}_2,\dots,\vec{x}_N]^T$
with their corresponding constant local coordinates
$\mathit{R}^0=[\vec{x^0}_1,\vec{x^0}_2,\dots,\vec{x^0}_N]^T$,
M pairs of collective variables
    $\vec{\theta}=[\psi_1,\psi_2,\cdots,\psi_M,\varphi_1,\varphi_2,\cdots,\varphi_M]^T$
, 
and the overall operator $\mathcal{P}$ being a combination of all the global operators $\hat{P}_I$.
The inverse transformation for obtaining all the angles is written as 
\begin{equation}
    \vec{\theta}=\hat{\Theta}(\mathit{R}).
\end{equation}
To overcome the periodicity of angles, in the network we project angles to sine-cosine values as
\begin{equation}
    \bold{S}=\mathbb{P}(\vec{\theta})=[\cos\psi_1,\cdots,\cos\psi_M,\sin\varphi_1,\cdots,\sin\varphi_M]^T.
\end{equation}
\par
The SDE is typically expressed as
\begin{equation}\label{eq:general-SDE}
\begin{split}
    \frac{dx(\tau)}{d\tau}&=f(x(\tau))+\sum_\alpha g_\alpha(x(\tau))\xi_\alpha
\end{split}
\end{equation}
where $x(\tau)$ represents the position within the system's phase or state space at time $\tau$. Here, $f$ is a flow vector field or drift force that signifies the deterministic evolution law, while $g_\alpha$ is a collection of vector fields that define the system's coupling to Gaussian white noise $\xi_\alpha$\cite{Slavik2013}.
\par
For the evolution of the CV vector $S$ at timestamps t and t+1, the SDE can be expressed as 
\begin{equation}\label{eq:cv-sde-simple}
    S_{t+1}
    =\mathbb{F}_0(S_t)+\mathbb{P}(\sigma^2_t\xi), \sigma_t\equiv \sigma\left(S(t)\right),
\end{equation}
where $\mathbb{F}_0$ represents the deterministic drift force component, and $\xi\equiv\xi(0,1)$ is the Gaussian noise.
Generalizing to $i\geq 1$ timestamps, the SDE from Eq.~\ref{eq:cv-sde-simple} can be written as
\begin{equation}\label{eq:sde-network-formula}
S_{t+i}=
\overset{i}{\overbrace{\mathbb{F}_0\circ\mathbb{F}_0\cdots\circ\mathbb{F}_0}}(S_{t})+\mathbb{P}(\sigma_{t,i}\xi),\sigma_{t,i}\equiv\sigma_{t,i}\left(S(t)\right)
\end{equation}
with $\sigma_{t,i}$ denoting the standard deviation dependent on the CVs at time $t$ and step size $i$.
It is important to note that the "addition" operator and noise term in Eq.~\ref{eq:cv-sde-simple} and ~\ref{eq:sde-network-formula} involve operations on angles, represented compactly using S.
Details of derivation of Eq.~\ref{eq:sde-network-formula} and~\ref{eq:cv-sde-simple} can be found in Sec.~\ref{sec:SDE-overview}
\par
We employ distinct neural networks, a propagator $\mathbb{F}$ for the drift force and a noise generator $\mathbb{G}$.
Eq.~\ref{eq:cv-sde-simple} for the noise component can be reformulated using these networks as follows:
\begin{equation}\label{eq:sde-simple-full}
    S_{t+1}=\mathbb{F}(S_t)+\mathbb{P}\circ\mathbb{G}^{-1}\left(\mathbb{F}(S_t),\xi\right),
\end{equation}
as illustrated in the upper part of Fig.~\ref{fig:network}.
The networks' task is now to predict the subsequent CVs based on the current ones, where these predicted variables are also projected to sine-cosine values.
In the following sections, we will delve into the drift force and noise components separately.
\subsubsection{Propagator for drift force}
\begin{figure}
\begin{subfigure}{0.5\textwidth}
\includegraphics[width=\textwidth]{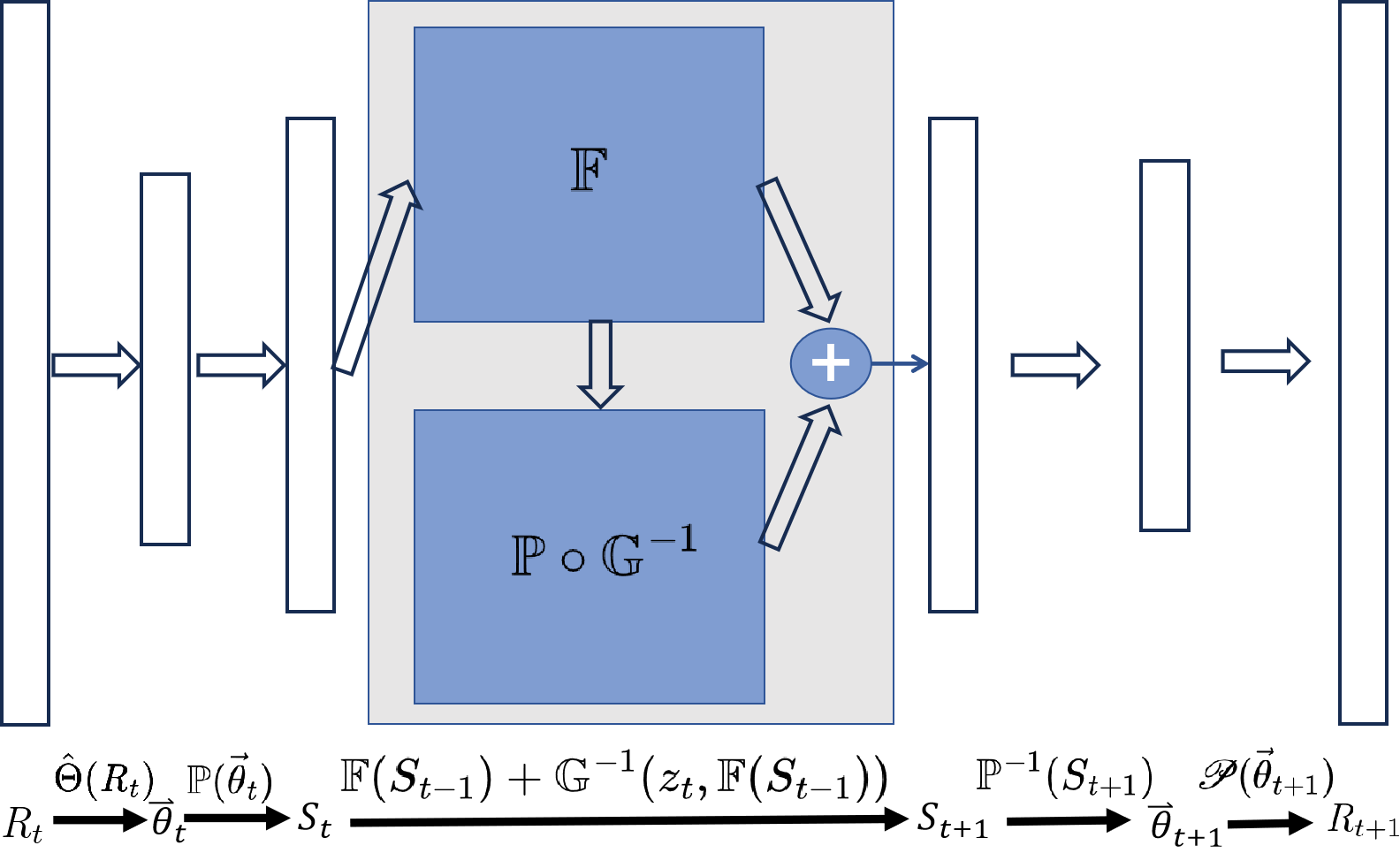}
\end{subfigure}
\begin{subfigure}{0.5\textwidth}
\includegraphics[width=\textwidth]{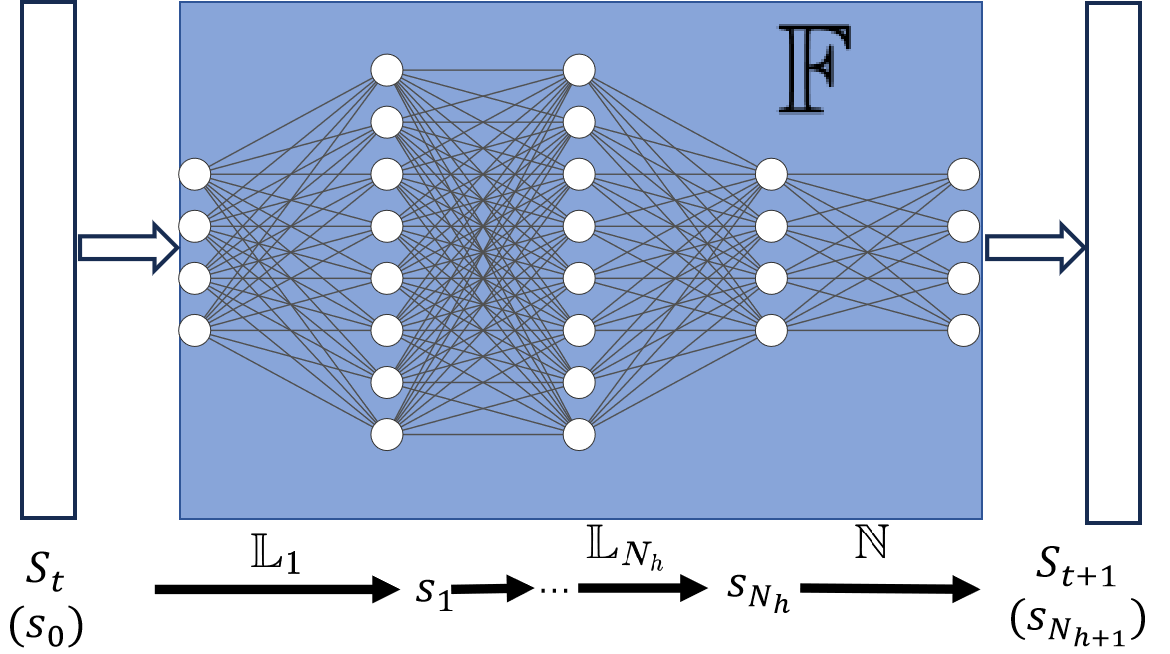}
\end{subfigure}
\begin{subfigure}{0.5\textwidth}
\includegraphics[width=\textwidth]{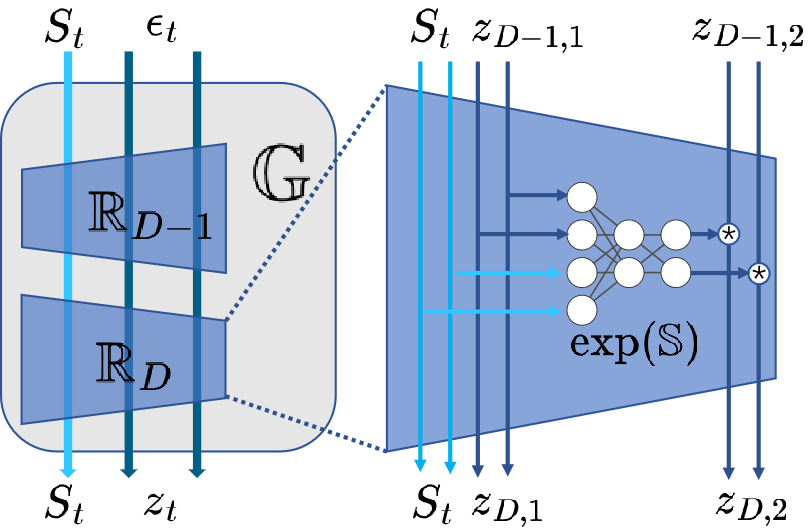}
\end{subfigure} 
\caption{
Deep neural network architecture for predicting the next protein coordinate. The upper panel illustrates the overall network structure, including modules for solving the SDE and coordinate transformation. The middle panel details the propagator network, 
$\mathbb{F}$, which models the drift force component. The lower panel shows the noise generator network, $\mathbb{G}$, responsible for generating the noise term.}
\label{fig:network}
\end{figure}

Without considering the noise term, Eq.~\ref{eq:sde-simple-full} degraded to the drift force propagator:
\begin{equation}\label{eq:all-network}
    \bold{S}_{t+1}= \mathbb{F}(\bold{S}_{t}) =\mathbb{N}\circ \mathbb{L}_{N_h}\circ\cdots\circ \mathbb{L}_{1}(\bold{S}_{t}),
\end{equation}
where $N_h$ denotes the number of hidden layers in the network, $t$ and $t+1$ represent the time indices (timestamps) of the trajectory data.
The structure of the network is illustrated in the middle part of Fig.~\ref{fig:network}.
The formula of function $\mathbb{L}_d$ in $d$th layer is written as
\begin{equation}\label{eq:linear-transform}
    \mathbf{s}_d=\mathbb{L}_d(\mathbf{s}_{d-1})=\mathbb{A}(\mathbf{W}_d\mathbf{s}_{d-1}+\mathbf{b}_{d-1}),
\end{equation}
where $\mathbf{W}_d\in \Re^{M_d\times M_{d-1}}$ and $\mathbf{b}\in \Re^{M_d}$ are weights to be learnt, and $\mathbb{A}$ are activation functions.
Our network distinguishes itself from Zhang et al.'s ~\cite{Zhang2018} work by introducing a normalization constraint, denoted by $\mathbb{N}$, for the sine-cosine operation within the network. 
This normalization step helps mitigate potential artifacts arising from the projection of angular data into sine and cosine components.
While these constraints could also be incorporated into the loss function, we found that applying them directly through $\mathbb{N}$ significantly accelerates the convergence of the network during training.
\par
The task is to obtain the weights $\left \{\mathbf{W}_d, \mathbf{b}\right\}$ that minimize the loss that can be described as $\min_{\left \{\mathbf{W},\mathbf{b}  \right\} }L_T$.
We use mean squared error (MSE) as the loss function which goes as
\begin{equation}\label{eq:loss-simple}
    L_T=\frac{\sum_{t=1}^{T-1} \left \Vert \mathbf{S}_{t+1}-\mathbb{F}(\bold{S}_{t})  \right \Vert^2}{T-1},
\end{equation}
where we use error between variables at $t+1$ and those obtained from $t$.
This loss can also be extended to a more general form as
\begin{equation}\label{eq:loss-general}
    L_{T_n}=\frac{1}{n}\sum_{i=1}^{n}\frac{1}{T-i} \sum_{t=1}^{T-i} \left \Vert \mathbf{S}_{t+i}- \overset{i}{\overbrace{\mathbb{F}\circ \cdots\circ \mathbb{F}}}  (\bold{S}_{t})  \right \Vert^2,\space 1\leq n\leq T-1,
\end{equation}
where the error computed from subsequent coordinates after $t+1$ are also used.
It is easy to see Eq.~\ref{eq:loss-general} collapses to Eq.~\ref{eq:loss-simple} with $n=1$.
\par
We will show Eq.~\ref{eq:loss-general} helps learns the true drift force of the SDE.
A subset of loss function in Eq.~\ref{eq:loss-general} for predicting the subsequent coordinate $S_{t+i}$ based on current time stamp $S_t$ for any time indexes $T_t=\left\{t_j|S_{t_j}=S_t,1\leq j\leq J\right\}$ can be written as
\begin{equation}\label{eq:loss-general-for-drift}
    \begin{split}
L^i_{T_t}&=\frac{1}{J}\sum_{j=1}^{J}\left \Vert S_{t+i}^j-S_{t+i,0}^j\right\Vert^2\\
        &=\left\Vert\mathbb{F}_0^i(S_t)-\mathbb{F}^i(S_t)\right\Vert^2+\left \Vert \sigma_{t,i}^2\right \Vert\\
    \mathbb{F}_0^i&\equiv \overset{i}{\overbrace{\mathbb{F}_0\circ\mathbb{F}_0\cdots\circ\mathbb{F}_0}},\mathbb{F}^i\equiv \overset{i}{\overbrace{\mathbb{F}\circ\mathbb{F}\cdots\circ\mathbb{F}}}
    \end{split},
\end{equation}
where $j$ indexes the training data, $i$ indicates the number of steps after the current time stamp and subscript $0$ refers to the true CVs.
The detailed derivation of this loss function can be found in Sec.~\ref{sec:SDE-drift} in the appendix.
It can be seen when the loss in Eq.~\ref{eq:loss-general} reaches its minimum, the trained DNN effectively approximates the drift force $\mathbb{F}_0$ represented in Eq.~\ref{eq:sde-network-formula}.
\subsubsection{RealNVP-based noise generator }\label{sed:noise-generator}
\par
RealNVP~\cite{Dinh2017} is a versatile method for estimating probability distributions.
Building upon this foundation, the Boltzmann generator\cite{Noe2019} was introduced to identify intermediate states in molecular dynamics simulations. 
Similarly, Timewarp\cite{Klein2023} leveraged RealNVP to construct a network for predicting noise terms in coordinate and velocity SDEs. 
Inspired by these works, we propose a simplified RealNVP-based noise generator. Given that our drift force component is modeled by a separate network, the residual noise is zero-mean, requiring only standard deviation estimation. Consequently, our noise generator architecture is more streamlined compared to Timewarp.
\par
Unlike the drift force calculation, we employ angles as collective variables for noise estimation. 
This choice is justified as we are concerned with distributions, where the periodicity of angles does not impact the computation. 
To generate noise data, the corresponding noise term in Eq.~\ref{eq:cv-sde-simple}, $\epsilon_t=\sigma_t\xi$  for angular noise can be derived from two consecutive collective coordinates as follows:
\begin{equation}
    \epsilon_t= \mathbb{P}^{-1}S_t-\mathbb{P}^{-1}\mathbb{F}(S_{t-1}),
\end{equation}
where $\mathbb{F}$ is the previously determined drift force.
\par
Consistent with the RealNVP framework, applying the transformation
$G$ to the noise term, $\epsilon_t$, yields a transformed variable, $z_t$ , that follows a normal distribution:  
\begin{equation}
    \mathbb{G}(\epsilon_t,S_t)=z_t,z_t\sim \mathit{N}(0,I).
\end{equation}
The operation network could be formulated by RealNVP layers $R$: 
\begin{equation}
\begin{split}
    \mathbb{G}=&\mathbb{R}_{N_h}\circ\cdots\circ \mathbb{R}_{1}\\
    z_D=&\mathbb{R}_D(z_{D-1},S_t),\space z_0=\epsilon_t,\space z_{N_h}=z_t
\end{split}.
\end{equation}
Each modified RealNVP layer $R_D,1\leq D\leq N_h$ maps:
\begin{equation}
    \mathbb{R_D}:\begin{bmatrix}
z_{D,1} \\z_{D,2}
\end{bmatrix}
= \begin{bmatrix}
z_{D-1,1}
\\
z_{D-1,2}\odot \exp \left(\mathbb{S}_D(z_{D-1,1},S_t)\right)
\end{bmatrix}
\end{equation}
\begin{equation}
    \mathbb{R_D}^{-1}:\begin{bmatrix}
z_{D-1,1} \\z_{D-1,2}
\end{bmatrix}
= \begin{bmatrix}
z_{D,1}
\\
z_{D,2}\odot \exp \left(-\mathbb{S}_D(z_{D,1},S_t)\right)
\end{bmatrix}
\end{equation}
\begin{equation}
    \mathbb{S}_D(z_{D-1,1},S_t)= \mathbb{L}_{N_D}\circ\cdots\circ \mathbb{L}_{1}
    \begin{bmatrix}
        \mathcal{P}^{-1}(S_t)
        \\
        z_{D-1,1}
    \end{bmatrix}
\end{equation}
where operator $\mathbb{L}_d(1\leq d\leq N_D)$  shares the form of Eq.~\ref{eq:linear-transform}.
Unlike the standard RealNVP architecture, our model incorporates only a scale term, $\mathbb{S}$, omitting the transformation term.
This simplification is justified by the zero-mean property of the noise term $\epsilon_t$, inherent to SDEs, and the inclusion of the coordinate component, $S_t$ within the scale term network. 
Excluding the input layer, the remaining network layers mirror those of $Z_{D-1,2}$.
To enhance the model's capacity, the positions of $z_{D,1}$ and $z_{D,2}$ are swapped after each RealNVP layer.
\par
For a given $z_t$, the probability of generated noise is
\begin{equation}
    p(\epsilon_t,S_t)=p(z_t)\exp\left(\sum_{D=1}^{N_h}\mathbb{S}_D(z_{D,1},S_t)\right).
\end{equation}
Network parameters are optimized by maximizing the acceptance ratio, defined as:
\begin{equation}
\begin{split}
    r(S_{t-1}, \tilde{S}_t)=&
    r(S'_t,\tilde{S}_t)=
    \frac{\mu(\tilde{S}_t)p(S'_t,\tilde{S}_t)}{
    \mu(S'_t)p(\tilde{S}_t,S'_t)}\\
    =&
    \frac{\mu(S'_t+\epsilon_t)p(-\epsilon_t,S'_t+\epsilon_t)}{
    \mu(S'_t)p(\epsilon_t,S'_t)
    }\\
    \tilde{S}_{t}=&\mathbb{F}(S_{t-1})+\epsilon_t=\mathbb{F}(S_{t-1})+\mathbb{P}\circ\mathbb{G}^{-1}z_t
    \\
    S'_t=&\mathbb{F}(S_{t-1})
\end{split},
\end{equation}
where $\mu(S_t)$ represents the probability of a given CV configuration. This probability is learned using a standard RealNVP network, as detailed in Dinh et al.'s work~\cite{Dinh2017}.
The loss function is 
\begin{equation}
\begin{split}
    L_{N}=&\frac{1}{T}\sum_{t=1}^{T}\sum_{z_t} \log r(S_{t-1}, \tilde{S}_t)=\frac{1}{T}\sum_{t=1}^{T}\sum_{z_t} \log r(S'_t, \tilde{S}_t)\\
         =&\frac{1}{T}
    \sum_{t=1}^{T}\sum_{z_t}\log\frac{\mu(S'_t+\mathbb{G}^{-1}z_t)p(-\mathbb{G}^{-1}z_t,S'_t+\epsilon_t)}{
    \mu(S'_t)p(\mathbb{G}^{-1}z_t,S'_t)
    }
\end{split}.
\end{equation}
The training process involves the simultaneous generation of random noise, $z_t$, and minimization of the loss function, $L_N$.
\subsubsection{Model quality assessment}\label{sec:model-quanlity}
\newversion{
Due to the inherent uncertainty in the timestep of the SDE, where saved trajectories become uncorrelated after an unknown number of intervals, a range of interval sizes $n\geq2$ needs to be considered in practical applications.
This choice of $n\geq2$ not only enhances the accuracy of the drift term over a larger timescale but also helps prevent the system from becoming trapped in local minima, when only the propagator operates on the SDE without the noise term for a larger interval $n$. Interestingly, when $n\geq2$ in Eq.~\ref{eq:loss-general}, the network demonstrably learns the underlying physical form of the drift force, leading to a more precise simulation, see Sec.~\ref{sec:forcefield} in the appendix.
Therefore, we recommend using $n\geq2$ in practice. We will further substantiate this recommendation with numerical evidence in the application section.
}

\par
We employed two key strategies to assess the quality of the generated MD trajectories. First, we analyzed the statistical performance of all the CVs, similar to many other studies. This involved plotting the error bars associated with these CVs. 
Second, we quantified the overall structural similarity between the generated and reference trajectories by calculating the RMSD difference ($\Delta \text{RMSD}(t)$) between the predicted and reference RMSD values at each time step (t)
\begin{equation}\label{eq:rmsd-error}
    \Delta \text{RMSD}(t)=\left | \text{RMSD}(t) - \text{RMSD}_{Ref}(t) \right |.
\end{equation}
The reference RMSD ($\text{RMSD}_{Ref}$) was obtained from the Cartesian coordinates.
\section{\label{sec:application-total}Application}
\subsection{\label{sec:application-structure}Protein Structure reconstruction}
To evaluate the effectiveness of our coarse-grained methods, we selected a challenging protein target: the 168-amino acid protein T1027 (sequence: KPTENNEDFNIVAVASNFATTDLDADRGKLPGKKLPL-EVLKEMEANARKAGCTRGCLICLSHIKCTPKMKKFIP-GRCHTYEGDKESAQGGIGEAIVDIPAIPRFKDLEPME-QFIAQVDLCVDCTTGCLKGLANVQCSDLLKKWLPQR-CATFASKIQGQVDKIKGAGGD) from the CASP14 prediction sets\cite{Moult2005,Moult1995}. 
This protein presents a computational challenge for traditional MD simulations due to its relatively long sequence and the presence of free loops. Successfully predicting the motion of this protein using our coarse-grained approach would provide a robust validation of our methods' ability to handle complex protein dynamics.
\par
Our coarse-grained representation utilizes 1624 parameters (described in Section~\ref{sec:methods}).
This includes 812 bond angles and 812 dihedral angles, both on the backbone and side-chains.
Compared to the 7665 parameters needed for a single protein's Cartesian coordinates and the significantly larger number required for full MD simulations (presented in Section~\ref{sec:application-md}), our approach offers a substantial reduction in dimensionality.
The total number of CVs is larger than
the representation using all dihedral angles for backbones that requires 504 parameters (3 angles/residue * 168 residues).
This is because we include the corresponding bond angles and related side chain information. This additional complexity allows for a more precise and bidirectional representation of the protein structure.
\par
\begin{figure}
\centering
\begin{subfigure}{0.42\textwidth}
\includegraphics{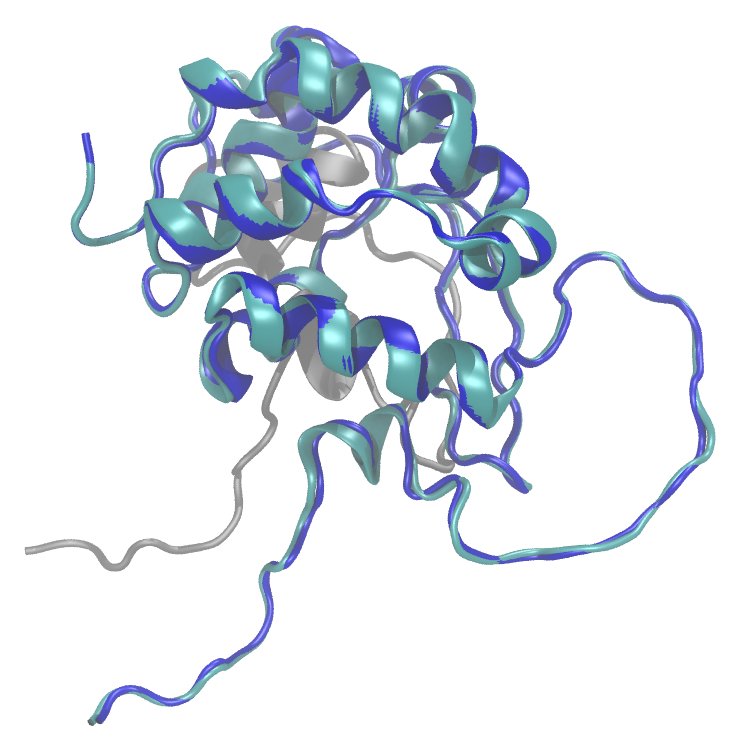}
\end{subfigure}
\begin{subfigure}{0.42\textwidth}  
\includegraphics{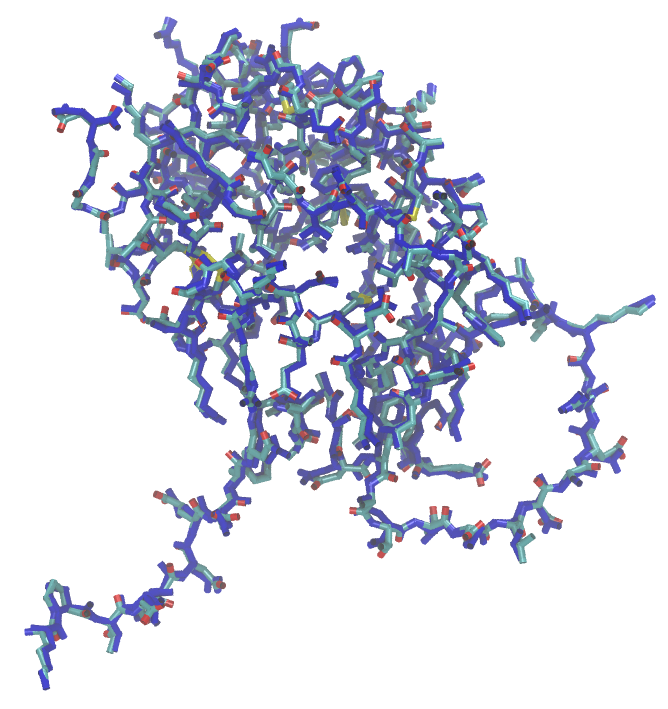}
\end{subfigure}
\caption{Cartesian coordinates reconstruction using dihedral and bond angles. The upper figure compares the tertiary structures and the lower one compares the secondary structures that show side-chains consistency. The original 3D structure is colored blue and the cyan structure is constructed by all real dihedral and bond angles. The gray tertiary structure on the upper figure is constructed by all the real angle data except $sp^3$ bond angles are fixed to 1.29 rads ($109^{\circ}28'$).}
\label{fig:T1027-illustration}
\end{figure}

Fig.~\ref{fig:T1027-illustration} depicts the reconstructed initial structure of protein T1027. \nycm{The cyan and blue structures both exist in the upper and lower figure, the difference is that the upper figure only shows the secondary structure of proteins and the lower figure shows more details, e.g. the sidechains. The gray structure is a bad one to show the importance of bond angles, and I assume it is better to show it later?} The reconstructed tertiary structure (cyan) exhibits a near-native match with the original data (blue), highlighting the accuracy of our approach. For comparison, the gray structure was built using all angles except for the $sp^3$ \ny{bond} angles, which were fixed at their theoretical value ($109^\circ28'$). This comparison reveals a significant mismatch, including the absence of part of the alpha helix. This demonstrates the substantial impact of even small variations in bond angles on protein structure, emphasizing the importance of including them in our representation.
The secondary structure representation also shows good agreement between the reconstructed structure and the original data, with only minor deviations observed at the ends of a few amino acid side chains. 
Specifically, analysis of CA atom distances reveals a maximum deviation of 0.26 angstroms, with an average value of 0.04 \ny{angstroms}\nycm{All "angstrom" are changed to "angstroms"}. The maximum difference for side-chain atoms is 0.6 angstroms, with a mean of 0.26 angstroms.
It is noteworthy that we utilize local coordinates (denoted as $x^0$ - see Section~\ref{sec:methods-cv}) for each amino acid, retrieved from a separate database independent of the T1027 structure. This approach can introduce slight error accumulation along the protein chain backbone, which can then propagate to the side chains (whose positions are highly dependent on the backbone). However, as will be shown in subsequent sections, the errors introduced by our coarse-grained representation remain consistently below 0.2 angstroms in practical applications, making them effectively negligible.
\subsection{\label{sec:application-md}MD trajectories prediction}
\par
We employed the GROMACS\cite{VanDerSpoel2005,Lindahl2001,Berendsen1995} software (version 2022.2) for the MD simulations, utilizing a timestep of 1 femtosecond (fs) and saving trajectories every 0.05 nanoseconds (ns). To enhance the efficiency of the simulations, we implemented an in-house developed enhanced sampling method, parallel continuous simulated tempering and a structure based model (PCST+SBM) within GROMACS.
\nyrd{
Using SBM focuses on the structural variations near a reference structure. This is because comprehensively analyzing all possible protein configurations is computationally demanding for large proteins, and this paper prioritizes a different aspect of the analysis.
}
For a detailed description of PCST+SBM, please refer to references~\cite{Zhang2010,Zang2014,Zang2018,Ma2018}. 
We start the MD simulations from the refinement target in CASP14.
Two copies are simulated in PCST, whose temperature distribution is defined by $\beta_0=0.38,\sigma_0=0.05$ and $\beta_1=0.27,\sigma_1=0.13$, respectively.
The replica exchange is performed every 5 ns and
the time step for updating temperature in Langevin equation is 0.01 ps.
In configuring the SBM method, we selected the starting target structure as the reference to promote initial-state fidelity during the MD simulations. This ensures the dynamics primarily focus on exploring configurations near the initial structure. Compared to the original SBM implementation, we introduce a targeted constraint strategy. We constrain only the "slow" residues, which are crucial for calculating the RMSDs. Residues within loops are left unconstrained to facilitate efficient exploration of the conformational space.
\par
For dataset preparation, we carefully select data points from the saved trajectories. Since the prediction of coordinates is inherently time-dependent, the data are sorted chronologically. Additionally, we address the artificial discontinuity introduced by replica exchange steps within the PCST method\cite{Zang2018}. These exchanges, motivated by temperature differences, are not representative of the actual physical system governed by the force field. Therefore, we re-arrange the trajectories to create a continuous data stream for training. This ensures the DNN learns the underlying dynamics without being confounded by non-physical artifacts.
\par
\ym{This coarse-grained MD simulation utilizes saved trajectories from a prior all-atom MD simulation. The all-atom simulation employed a timestep of 1 fs with data saved every 0.05 ns for a total simulation length of 250 ns.
Since this study focuses on replicating the time-series behavior of the simulation, utilizing the full configuration space data becomes unnecessary.  While SBM inherently doesn't provide access to this complete data, obtaining it for such a long protein would also be computationally expensive. Our focus on time-series allows us to achieve a more precise result without requiring the full configuration space.}
For training the deep neural network, we employ a common data split strategy: 70\% of the processed trajectories are used for training, 20\% for validation, and the remaining 10\% for testing.
We configure the network architecture with three hidden layers ($N_h=3$) and utilize two weight matrices ($M_d$) of size $3248\times3807$ and $3807\times3248$ (details on weight matrix dimensions might be specific to your field).
\subsubsection{Drift force only}
\begin{figure}
\centering
\begin{subfigure}{0.5\textwidth}
    \includegraphics[width=\textwidth]{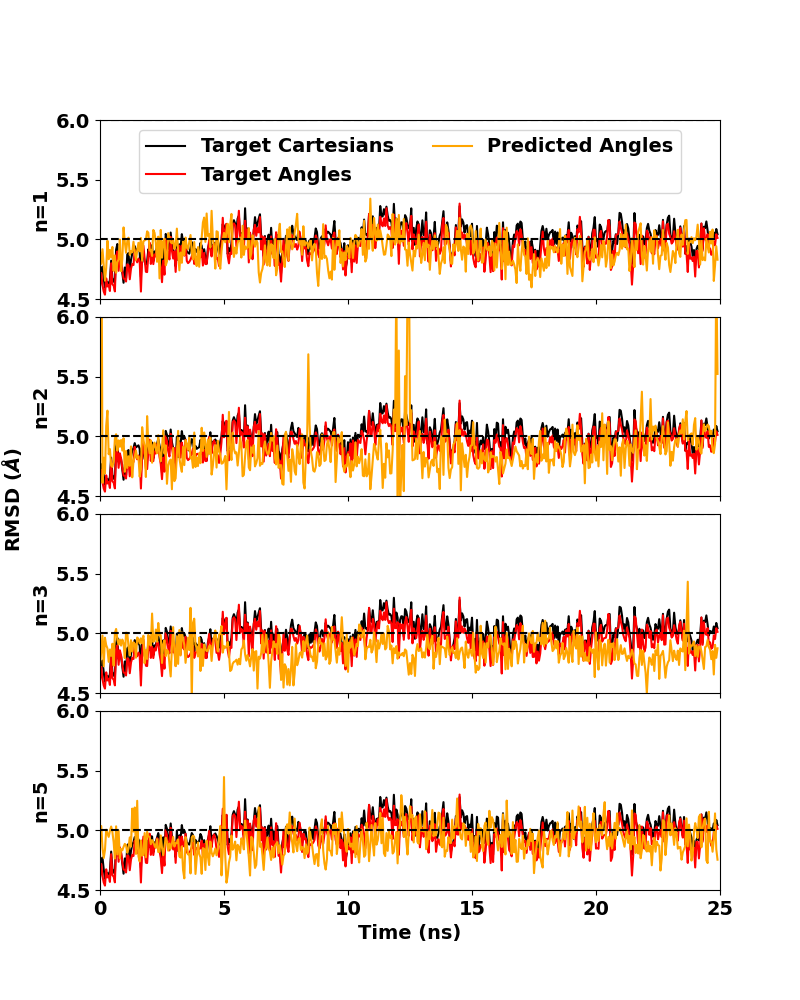}
    \caption{RMSDs for n = 1, 2, 3 and 5.}
\end{subfigure}
\begin{subfigure}{0.45\textwidth}
    \includegraphics[width=\textwidth]{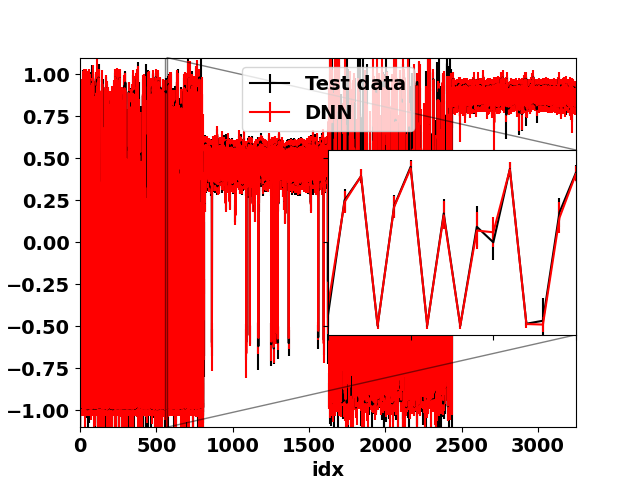}
    \caption{$\vec{\theta}$ statistics with n = 1.}
\end{subfigure}
\caption{RMSD of CA atoms (upper) and statistical performance of angles (lower) in the test datasets. 
Readers are redirected to CASP14 website for RMSD calculation methods. In the lower figure, the values in the nodes are the mean values and the vertical lines in the node show the standard deviations.
In the upper figure, the black line represents rmsds calculated from original trajectories, the red line represents the rmsds obtained from structures re-constructed from CVs and the orange line represents the rmsds from predicted structures.
}
\label{fig:MD-results}
\end{figure}
We first do simulations only with drift force propagator using Eq.~\ref{eq:all-network} to show the accuracy of predicted MD trajectories using the test dataset.
Fig.~\ref{fig:MD-results} depicts the predicted performance of the MD simulation. The upper panel shows the angles (torsion angles $\psi$ and bond angles $\pi-\varphi$).
As expected, the $\psi$ angles (represented in the first half) exhibit a wider distribution range due to their inherent flexibility in the protein backbone. In contrast, the $\varphi$ angles ($\pi-\varphi$, shown in the second half) primarily cluster around 1.23 radians ($109^{\circ}28'$) with a smaller variance (~0.4 radians) during the simulation.
This finding reinforces the earlier observation of bond angle fluctuations within MD simulations.
Notably, a small population of $\varphi$ angles around $\frac{2\pi}{3}$ radians likely originates from the carbonyl oxygen atoms in the backbone.
The comparison of the orange (black) and red (predicted) curves reveals a high degree of consistency between the predicted statistical behavior of the angles and the original values. Similar observations can be made when analyzing the sine and cosine components of these angles, as presented in Fig.~\ref{fig:sincos-error} in the appendix.
\par
The RMSD values calculated from the predicted CVs closely match those obtained from the original Cartesian coordinates (Fig.~\ref{fig:MD-results}, black vs. red lines).
Our observations align well with previous structural analysis, where deviations were consistently less than 0.2 angstroms and the average difference using angles ($\Delta \text{RMSD}_{Angle}(t)$) was 0.0691 angstroms with a standard error of 0.0418 angstroms.
Additionally, the predicted trajectories closely resemble the original data for all four trajectories, particularly between 0-5 ns and 10-20 ns for $n=1$  and $n=5$, as shown by the RMSD profiles.
\par
A convergence study (Fig.
~\ref{fig:rmsd-convergence}, Appendix) was conducted to evaluate the performance of our method for different values of n (1, 2, 3, and 5). Numerical error was calculated using Eq.~\ref{eq:rmsd-error}.
While all four n values exhibited comparable performance in short (25 ns) simulations (black curve), longer simulations (up to 175 ns) demonstrated that $n\geq2$ outperformed $n=1$, with convergence achieved at $n=2$ (red line). This suggests that a minimum of two iterations is necessary to capture the essential force field information and generate accurate MD simulations.
A more in-depth analysis (Sec.~\ref{sec:forcefield}, Appendix) indicates that only $n\geq 2$ effectively learns relevant force information. However, further investigation is required to provide a conclusive proof and a systematic convergence study.
For practical applications, we found that directly using the drift force alone was not optimal. Combining $n=1$ with the relevant noise component proved sufficient for long simulations, providing a balance between computational efficiency and accuracy.
\par
Our DNN exhibits remarkable efficiency. 
The propagation process of a 25ns trajectory using only the initial structure as input takes merely 0.59 seconds on a standard NVIDIA GeForce GTX 1660 Ti (6G) GPU. 
In stark contrast, traditional MD simulations on a 256-CPU supercomputer require roughly 20 hours to generate a comparable trajectory. 
This translates to an impressive speedup of approximately 70,000 times achieved by the DNN network.
Even when considering the training time (around 2 hours), the coarse-grained DNN network offers a significant speedup factor of 10 for the entire process. 
It's important to note that training can be further optimized by leveraging parallelization techniques and better GPUs.
\subsubsection{Drift force and noise}
\begin{figure}
\centering
    \includegraphics[width=0.45\textwidth]{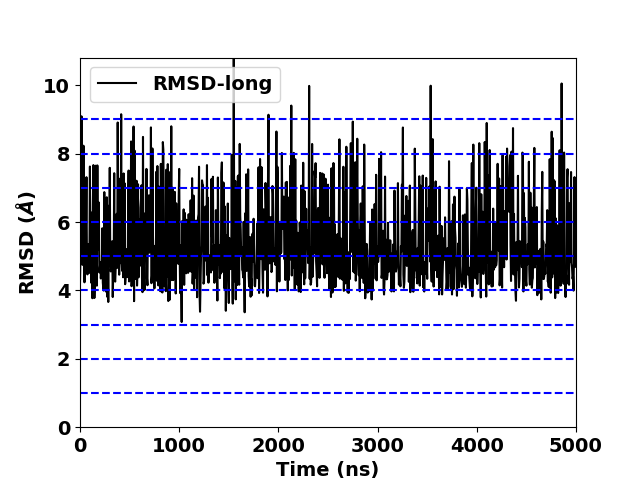}
\caption{Simulation with first frame of the testing data. Noise coefficient is and a constant temperature simulation is performed. Its lowest RMSD is around 3.08 \AA, lower than the lower limit 3.83 \AA from the training data.}
\label{fig:long-simulation}
\end{figure}
Subsequent simulations employed a pre-trained propagator combined with noise terms generated by the modified RealNVP-based noise generator described in Section~\ref{sed:noise-generator}. Generating a 5000 ns trajectory required approximately 10 minutes of computational time, a significant acceleration of 24,000 times compared to traditional MD simulations, which would take around 4000 hours. 
Considering the noise generator training time of approximately 1 hour, the overall speedup is estimated at around 100 times.
\par
Figure~\ref{fig:long-simulation} presents the resulting RMSD trajectories, which exhibit relatively large and variable RMSD values compared to the test data. 
This variability can be attributed to the stochastic nature of the generated noise and potential numerical errors arising from configurations that extend beyond the recorded configuration space due to the limitations of SBM in fully encompassing the conformational space. 
Interestingly, a configuration with an RMSD of 3.08 Å was sampled, demonstrating closer proximity to a local minimum than any configuration within the training trajectory. 
This observation highlights the potential for our method to function as a standalone molecular dynamics simulation tool.
\par
\begin{figure}
\includegraphics[width=0.45\textwidth]{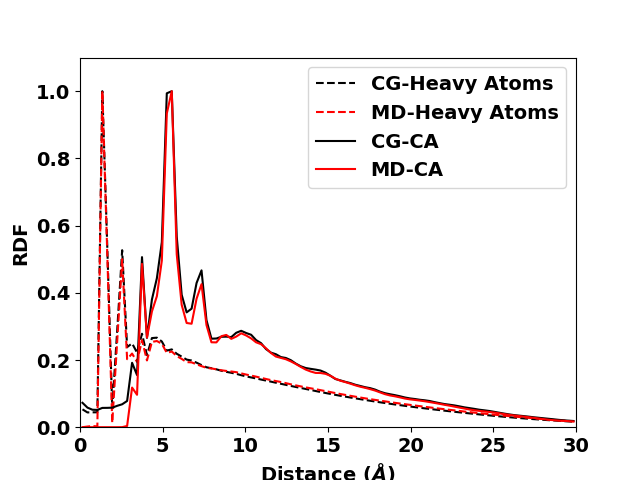}
\caption{The RDFs of CA atoms (solid lines) with those of all heavy atoms (dashed lines) for both the atomistic MD simulation (red line) and the generated CG trajectories (black line).
All RDFs are normalized to a maximum value of 1 for better comparison. Importantly, only intra-protein atom distances are considered in the RDF calculations. Inter-protein atom distances are excluded.}
\label{fig:RDF}
\end{figure}

To enhance our trajectory analysis,
we conducted a statistical assessment across ten additional trajectories, accumulating 250 ns of simulation data. We calculated radial distribution functions (RDFs) for both all heavy atoms and the CA atoms employed for RMSD calculations (Fig.~\ref{fig:RDF}). The RDF for all heavy atoms reveals a prominent peak at 1.35 Å, indicative of the average atomic bond length. Importantly, the maximal RDF discrepancies between coarse-grained (CG) and molecular dynamics (MD) simulations are minimal (0.053 for all heavy atoms and 0.074 for CA atoms, normalized to 1), demonstrating the high fidelity of our CG model in reproducing MD trajectory RDFs.

\par
Our CG method achieves a more precise representation of protein dynamics, as evidenced by the close agreement between our predicted and reference RMSD values, which was never achieved in previous CG methods\cite{Wang2022,Zhang2018,Yu2021,Wang2019,Husic2020}.
The superior precision stems from two key features.  First, incorporating both bond and dihedral angles allows for near-native reconstruction of the protein backbone, significantly improving its structural representation. Second, including side-chain information enhances the precision of interactions between amino acids, leading to more realistic protein behavior.
Furthermore, the ability to perform bi-directional conversion between CG representation and all-atom structure empower our CG simulations to function as a truly independent tool, extending beyond a complementary role to traditional all-atom simulations\cite{Wang2022,Zhang2018}.
\section{\label{sec:conclusion}Conclusion and outlook}
This research introduces a novel method using artificial intelligence to perform rapid, coarse-grained simulations of protein molecules. Our approach establishes a two-way link between simplified protein descriptions (CVs) and their detailed 3D structures. This link, built using tree-like coefficients, allows for accurate reconstruction of protein structures.
Using a DNN, we successfully generated the MD trajectories of a 168-amino acid target protein (T1027) on a standard computer. This accomplishment is remarkable because it achieves speeds 70,000 times faster than traditional MD simulations, which require powerful supercomputers. Importantly, the generated trajectories closely match the statistical behavior and structural properties observed in real MD simulations.
By employing a DNN and a dedicated noise generator, we can generate trajectories on the order of microseconds, achieving a speedup of over 20,000 times compared to traditional methods.
This work paves the way for highly efficient and accurate coarse-grained protein simulations using artificial intelligence.
\par
These future research directions hold significant promise for further refining and expanding the capabilities of the presented framework. 
\begin{enumerate}
\item \textbf{Enhanced Sampling within the Coarse-Grained Framework}.
While our training data incorporates enhanced sampling methods, the network may not fully capture the temperature dependence of the CV space. Future work will explore incorporating temperature and energy information directly into the DNN training process.
\item \textbf{Exploration of Alternative Training Methods}.
Graphical Neural Networks (GNNs) offer an intriguing possibility for incorporating the inherent hierarchical structure of proteins into the training process. GNNs may be particularly well-suited for capturing the differential contributions of backbone and side-chain interactions to the overall dynamics.
\item \textbf{Expanding Applicability to Other Systems}
The proposed framework has the potential to be extended beyond protein simulations. By utilizing appropriate CV representations, the method could be applied to MD simulations of Deoxyribonucleic acid (DNA), Ribonucleic acid (RNA), and even non-biochemical systems such as polymers and electrolytes in batteries.
\end{enumerate}

% \begin{acknowledgments}
% We are grateful for fruitful discussions with Dr. Ningyi Lv and Dr. Yaming Yan from Fudan University.
% \end{acknowledgments}

\appendix
\gdef\thefigure{\thesection.\arabic{figure}}    
\section{Example amino structure}
\setcounter{figure}{0}
\begin{figure}
\centering
    \includegraphics[width=0.45\textwidth]{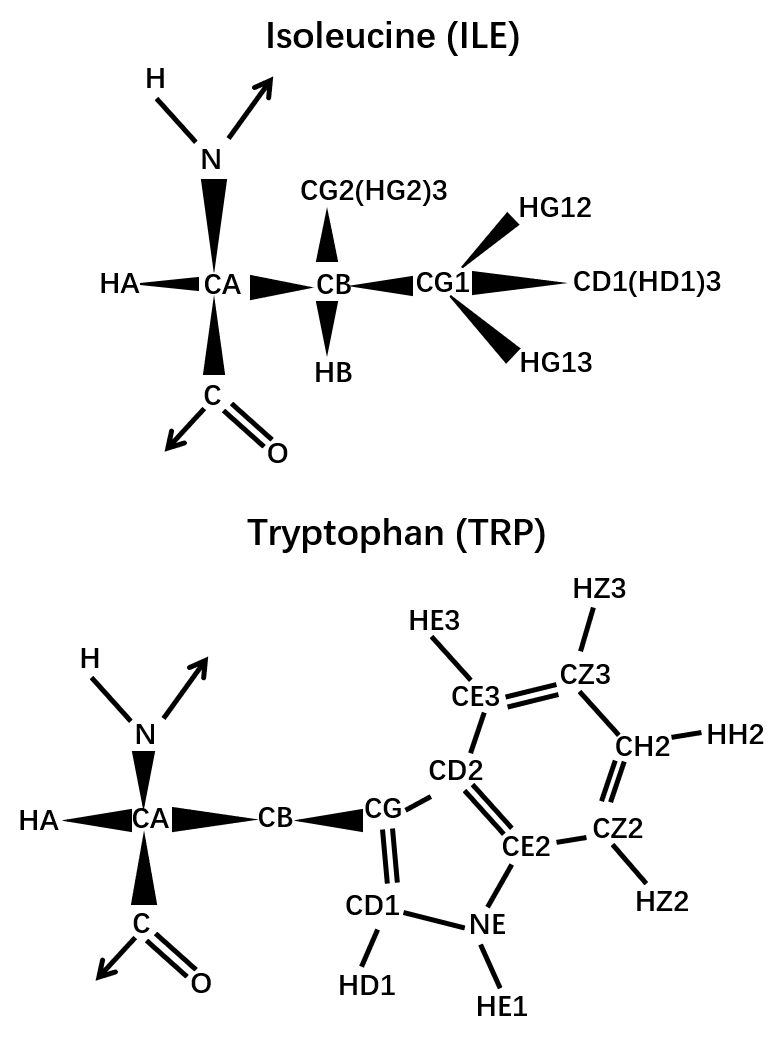}
    % \includegraphics[width=0.45\textwidth]{figures/TRP.png}
    % \caption{Firts subfigure.}
\caption{The illustration of structures of ILE (upper panel) and TRP (lower panel).}
\label{fig:ile-trp}
\end{figure}
\section{Amino hierarchy}
\setcounter{figure}{0}
% \usepackage{multirow}
\begin{table}[]
\caption{
Protein Structure Hierarchy (Ignoring Common N Atoms).
This figure depicts the hierarchical representation of a protein structure, with each box representing a group of atoms sharing a local coordinate frame.
The hierarchy starts with CA atoms in the first column, and progressively incorporates neighboring atoms in subsequent columns (increasing depth, show as $0,1,2\cdots 6$) in the first row.}
\label{tab:amino-hierarchy}
\begin{tabular}{lccccccc}
\hline
Amino                & 0                   & 1                   & 2       & 3                                                                             & 4                                                 & 5  & 6       \\ \hline
ALA                  & CA                  & CB                  &         &                                                                               &                                                   &    &         \\ \hline
                     &                     & C                   & O       &                                                                               &                                                   &    &         \\ \hline
\multirow{2}{*}{VAL} & \multirow{2}{*}{CA} & CB                  & CG1,CG2 &                                                                               &                                                   &    &         \\ \cline{3-8} 
                     &                     & C                   & O       &                                                                               &                                                   &    &         \\ \hline
\multirow{2}{*}{LEU} & \multirow{2}{*}{CA} & CB                  & CG      & CD1,CD2                                                                       &                                                   &    &         \\ \cline{3-8} 
                     &                     & C                   & O       &                                                                               &                                                   &    &         \\ \hline
GLY                  & CA                  & C                   & O       &                                                                               &                                                   &    &         \\ \hline
\multirow{3}{*}{ILE} & \multirow{3}{*}{CA} & \multirow{2}{*}{CB} & CG1     &                                                                               &                                                   &    &         \\ \cline{4-8} 
                     &                     &                     & CG2     & CD                                                                            &                                                   &    &         \\ \cline{3-8} 
                     &                     & C                   & O       &                                                                               &                                                   &    &         \\ \hline
\multirow{2}{*}{MET} & \multirow{2}{*}{CA} & CB                  & CG      & SD                                                                            & CE                                                &    &         \\ \cline{3-8} 
                     &                     & C                   & O       &                                                                               &                                                   &    &         \\ \hline
\multirow{2}{*}{TRP} & \multirow{2}{*}{CA} & CB                        & \begin{tabular}[c]{@{}c@{}}CG,CD1,CD2\\ NE1,CE2\\ CE3,CZ2\\ CZ3,CH2\end{tabular}& &                                                   &    &         \\ \cline{3-8} 
                     &                     & C                   & O       &                                                                               &                                                   &    &         \\ \hline
\multirow{2}{*}{PHE} & \multirow{2}{*}{CA} & CB                  & CG      & \begin{tabular}[c]{@{}c@{}}CD1,CD2\\ CE1,CE2\\ CZ\end{tabular}                &                                                   &    &         \\ \cline{3-8} 
                     &                     & C                   & O       &                                                                               &                                                   &    &         \\ \hline
\multirow{2}{*}{PRO} & \multirow{2}{*}{CA} & CB                  & CG,CD   &                                                                               &                                                   &    &         \\ \cline{3-8} 
                     &                     & C                   & O       &                                                                               &                                                   &    &         \\ \hline
\multirow{2}{*}{SER} & \multirow{2}{*}{CA} & CB                  & OG      &                                                                               &                                                   &    &         \\ \cline{3-8} 
                     &                     & C                   & O       &                                                                               &                                                   &    &         \\ \hline
\multirow{2}{*}{CYS} & \multirow{2}{*}{CA} & CB                  & SG      &                                                                               &                                                   &    &         \\ \cline{3-8} 
                     &                     & C                   & O       &                                                                               &                                                   &    &         \\ \hline
\multirow{2}{*}{ASN} & \multirow{2}{*}{CA} & CB                  & CG      & OD1,ND2                                                                       &                                                   &    &         \\ \cline{3-8} 
                     &                     & C                   & O       &                                                                               &                                                   &    &         \\ \hline
\multirow{2}{*}{GLN} & \multirow{2}{*}{CA} & CB                  & CG      & CD                                                                            & OE1,OE2                                           &    &         \\ \cline{3-8} 
                     &                     & C                   & O       &                                                                               &                                                   &    &         \\ \hline
\multirow{2}{*}{THR} & \multirow{2}{*}{CA} & CB                  & OG1,CG2 &                                                                               &                                                   &    &         \\ \cline{3-8} 
                     &                     & C                   & O       &                                                                               &                                                   &    &         \\ \hline
\multirow{2}{*}{TYR} & \multirow{2}{*}{CA} & CB                  & CG      & \begin{tabular}[c]{@{}c@{}}CD1,CE1\\ CZ,OH\\ CE2,CD2\end{tabular}             &                                                   &    &         \\ \cline{3-8} 
                     &                     & C                   & O       &                                                                               &                                                   &    &         \\ \hline
\multirow{2}{*}{ASP} & \multirow{2}{*}{CA} & CB                  & CG      & \begin{tabular}[c]{@{}c@{}}OD1\\ OD2\end{tabular}                             &                                                   &    &         \\ \cline{3-8} 
                     &                     & C                   & O       &                                                                               &                                                   &    &         \\ \hline
\multirow{2}{*}{GLU} & \multirow{2}{*}{CA} & CB                  & CG      & CD                                                                            & \begin{tabular}[c]{@{}c@{}}OE1\\ OE2\end{tabular} &    &         \\ \cline{3-8} 
                     &                     & C                   & O       &                                                                               &                                                   &    &         \\ \hline
\multirow{2}{*}{LYS} & \multirow{2}{*}{CA} & CB                  & CG      & CD                                                                            & CE                                                & NZ &         \\ \cline{3-8} 
                     &                     & C                   & O       &                                                                               &                                                   &    &         \\ \hline
\multirow{2}{*}{ARG} & \multirow{2}{*}{CA} & CB                  & CG      & CD                                                                            & NE                                                & CZ & NH1,NH2 \\ \cline{3-8} 
                     &                     & C                   & O       &                                                                               &                                                   &    &         \\ \hline
\multirow{2}{*}{HIS} & \multirow{2}{*}{CA} & CB                  & CG      & \begin{tabular}[c]{@{}c@{}}ND1,CD2\\ CE1,NE2\end{tabular}                     &                                                   &    &         \\ \cline{3-8} 
                     &                     & C                   & O       &                                                                               &                                                   &    &         \\ \hline
                     &                     &                     &         &                                                                               &                                                   &    &        
\end{tabular}
\end{table}
\section{SDE with CVs}\label{sec:SDE}
In this section, we derive the relevant formulas introduced in Sec.~\ref{sec:SDE-CV}.
To simplify notation, we consistently employ $\mathbb{F}$ to represent the drift force function regardless of input variables.
\subsubsection{SDE overview}\label{sec:SDE-overview}
The noise term in Eq.~\ref{eq:general-SDE} can typically be expressed as a single Gaussian noise term.
The equation can be reformulated as follows:
\begin{equation}\label{eq:general-SDE-transform}
\begin{split}
    \frac{dx(\tau)}{d\tau}&=f(x(\tau))+\sum_\alpha g_\alpha(x(\tau))\xi_\alpha\\
    &= f(x(\tau))+\xi(0,\sigma^2(x(\tau)))\\
    \sigma(x(\tau))&\equiv \sqrt{\sum_\alpha g_\alpha^2 (x(\tau))}
\end{split},
\end{equation}
where $\xi(\mu,\sigma^2(x(\tau)))$ is a random variable drawn from a Gaussian distribution with mean $\mu$ and variance $\sigma^2(x(\tau))$.
Discretizing time with a constant interval of $\Delta \tau$, the position at the subsequent time step can be expressed as 
\begin{equation}
\begin{split}
    x(\tau+\Delta \tau)
    &=x(\tau)+f(x(\tau))\Delta \tau+\Delta \tau\xi(0,\sigma^2(x(\tau)))\\
    &=F_0(x(\tau))+\xi(0,(\sigma(x(\tau))\Delta \tau)^2)\\
    F_0(x)&\equiv x + f(x)\Delta \tau
\end{split}.
\end{equation}
Here, $\mathbb{F}_0$ denotes the true drift force component,
 while $\xi(0,\sigma^2(x(\tau))$ represents the Gaussian noise.
It can be written as
\begin{equation}\label{eq:angle-sde-simple}
    \vec{\theta}_{t+1}=\mathbb{F}_0(\vec{\theta}_t)+\xi(0,\sigma(\vec{\theta}_{t})).
\end{equation}
for our representations using angles $\theta$.
Replacing $\vec{\theta}_t$ with $S_t$ for notational brevity, Eq.~\ref{eq:angle-sde-simple} becomes Eq.~\ref{eq:cv-sde-simple}.
\par
Similarly, for the subsequent two steps can be expressed as
\begin{equation}
\begin{split}
    \vec{\theta}_{t+2}&=\mathbb{F}_0(\vec{\theta}_{t+1})+\xi(0,\sigma^2_{t+1})\\
    &=
\mathbb{F}_0\left(\mathbb{F}_0(\vec{\theta}_{t})+\xi(0,\sigma^2_{t})\right)+\xi(0,\sigma^2_{t+1})\\
&\approx 
\mathbb{F}_0\circ\mathbb{F}_0(\vec{\theta}_{t})+\mathbb{F}_0^{\prime}(\mathbb{F}_0(\vec{\theta}_{t}))\xi(0,\sigma^2_t)+\xi\left(0,\sigma^2_{t+1}\right)\\
&=\mathbb{F}_0\circ\mathbb{F}_0(\vec{\theta}_{t})+\xi(0,\sigma^2_{t,2})\\
&\sigma_{t,2}\equiv\sqrt{\left[\mathbb{F}_0^{\prime}(\mathbb{F}_0(\vec{\theta}_{t}))\right]^2\sigma^2_{t} + \sigma^2_{t+1}}
\end{split}.
\end{equation}
Further derivation demonstrates that the coordinates at any arbitrary time step, denoted as $t+i$, can be expressed as the drift force originating from $t$ and a zero-mean Gaussian noise:
\begin{equation}\label{eq:sde-drif-force-angle}
\begin{split}
\vec{\theta}_{t+i}&=
\overset{i}{\overbrace{\mathbb{F}_0\circ\mathbb{F}_0\cdots\circ\mathbb{F}_0}}(\vec{\theta}_{t})+\xi\left(0,\sigma^2_{t,i}\right)\\
&=\mathbb{F}_0^i(\vec{\theta}_t)+\xi\left(0,\sigma^2_{t,i}\right)\\
\mathbb{F}_0^i&\equiv \overset{i}{\overbrace{\mathbb{F}_0\circ\mathbb{F}_0\cdots\circ\mathbb{F}_0}}
\end{split}.
\end{equation}
Using $S_t$ for notional brevity, Eq.~\ref{eq:sde-drif-force-angle} becomes Eq.~\ref{eq:sde-network-formula}.
\subsubsection{Drift force}\label{sec:SDE-drift}
We will demonstrate the proof for Eq.~\ref{eq:loss-general-for-drift} using a small data subset.
Initially, we will establish the equivalence of the MSE of $S$ and $\vec{\theta}$ within the loss function.
For two given $S$ values which are very close, $S_\alpha=[\cos\theta_\alpha, \sin\theta_\alpha]$ and $S_\beta=[\cos\theta_\beta, \sin\theta_\beta]$
\begin{equation}
\begin{split}
    &\left \Vert S_\alpha - S_\beta \right \Vert^2\\
    =&(\cos\theta_\beta-\cos\theta_\alpha)^2+(\sin\theta_\beta-\sin\theta_\alpha)^2\\
    =&1-2\cos(\theta_\alpha-\theta_\beta)
    =4\sin^2(\frac{\theta_\alpha-\theta_\beta}{2})\\
    \sim& (\theta_\alpha-\theta_\beta)^2.
\end{split}
\end{equation}
The MSE, used for predicting the subsequent coordinate $S_{t+i}$ based on the current time stamp $t$ for any time indexes $T_t=\left\{t_j|S_{t_j}=S_t,1\leq j\leq J\right\}$, can be expressed as
\begin{equation}\label{eq:loss-part}
    \begin{split}
L^i_{T_t}&=\frac{1}{J}\sum_{j=1}^{J}\left \Vert S^j_{t+i}-S^{j}_{t+i,0}\right\Vert^2\sim \frac{1}{J}\sum_{j=1}^{J}\left \Vert \vec{\theta}^j_{t+1} -\vec{\theta}^j_{t+1,0}\right\Vert\\
        &=\frac{1}{J}\sum_{j=1}^{J}\left\Vert\mathbb{F}_0^i(\vec{\theta}_{t})+\xi_j\left(0,\sigma_{t,i}^2\right)-\mathbb{F}^i(\vec{\theta}_{t})\right\Vert^2\\
        &=\left\Vert\mathbb{F}_0^i(\vec{\theta}_t)-\mathbb{F}^i(\vec{\theta}_t)\right\Vert^2+\frac{1}{J}\sum \left\Vert\xi_j\left(0,\sigma_{t,i}^2\right)\right\Vert^2\\
        &=\left\Vert\mathbb{F}_0^i(\vec{\theta}_t)-\mathbb{F}^i(\vec{\theta}_t)\right\Vert^2+\frac{1}{J}\sum \left\Vert\sigma_{t,i}\xi(0,1)\right\Vert^2\\
        &\approx\left\Vert\mathbb{F}_0^i(\vec{\theta}_t)-\mathbb{F}^i(\vec{\theta}_t)\right\Vert^2+\left \Vert \sigma_{t,i} \right \Vert^2\\
        &\approx \left\Vert\mathbb{F}_0^i(S_t)-\mathbb{F}^i(S_t)\right\Vert^2+\left \Vert \sigma_{t,i} \right \Vert^2\\
    \mathbb{F}_0^i&\equiv \overset{i}{\overbrace{\mathbb{F}_0\circ\mathbb{F}_0\cdots\circ\mathbb{F}_0}},\mathbb{F}^i\equiv \overset{i}{\overbrace{\mathbb{F}\circ\mathbb{F}\cdots\circ\mathbb{F}}}
    \end{split},
\end{equation}
where symbols $j$ and $i$ represent the index of the training data and the steps following the current timestamp, respectively. 
\section{Force-field information}\label{sec:forcefield}
We show here, when $n\geq2$ in Eq.~\ref{eq:loss-general}, the network effectively learns the underlying physical form of the force of the drift, which gives a more precise simulation.
We illustrate this by the velocity-verlet algorithm, which is a popular method in molecular simulations and allows for efficient calculation of forces acting on particles.
This method uses the particles' positions at three time steps (t-1, t, and t+1) as detailed in
\begin{equation}\label{eq:forcefield}
\begin{split}
     \bold{F}(\bold{S}_t)=&\bold{M}_S\times\bold{a}\approx\bold{M_S}\times\frac{\bold{S}_{t+1}+\bold{S}_{t-1}-2\bold{S}_t}{\Delta t^2}\\
     =&\bold{M_S}\times\frac{
     \mathbb{F}\circ\mathbb{F}(\bold{S}_{t-1})
     +\bold{S}_{t-1}-
     2\mathbb{F}(\bold{S}_{t-1})}{\Delta t^2}
\end{split}.
\end{equation}
The calculation works regardless of the particles' initial velocities. In Eq.~\ref{eq:forcefield}, $\times$ is the piece-wise multiplication operator, $\bold{M}_S$ represents a constant mass assigned to each simplified unit (coarse-grained coordinate), $\Delta t$ is the predefined time step, and the acceleration vector $\bold{a}$ is the only unknown variable to be solved for.
We would like to note that Eq.~\ref{eq:forcefield} is only an illustration to show the inclusion of force-field information with $n\geq2$ where many details are excluded.
For example, the much larger time step $\Delta t$ than the MD simulation indicates the computation of force is not accurate in the equation and the dependence of the velocity of the time due to our PCST method indicates the force should be dependent of time.
However, this equation helps indicate that least three successive trajectories enables the network to learn the drift force information.
\section{Propagation simulation}
\setcounter{figure}{0}
\begin{figure}
\centering
    \includegraphics[width=0.45\textwidth]{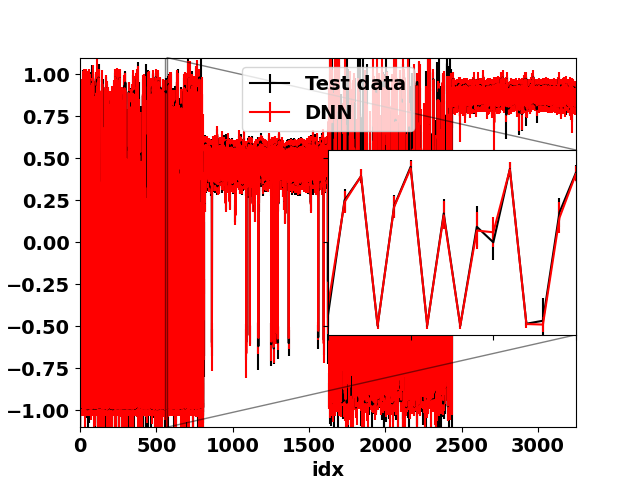}
    % \caption{Firts subfigure.}
\caption{The error of sine-cosine (\textbf{S}) of angles in MD simulation.}
\label{fig:sincos-error}
\end{figure}
\begin{figure}
\begin{subfigure}{0.42\textwidth}
    \includegraphics[width=\textwidth]{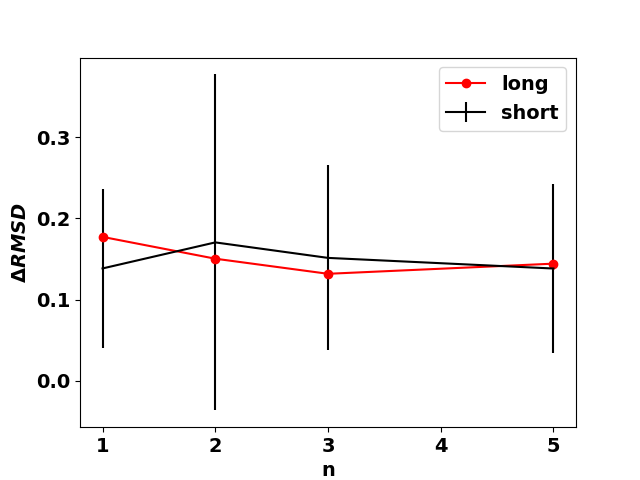}
    % \caption{Convergence Profile.}
\end{subfigure}
\caption{Convergence of RMSD Profiles for n = 1, 2, 3 and 5. Black Line: Error bars from short simulations on test data. Red Line: Mean error calculated from long simulations.}
\label{fig:rmsd-convergence}
\end{figure}
% angle mean 0.0691 error 0.0418, 
% step 1 mean 0.2457 error 0.2186
% step 2 mean 0.1285 error 0.0981
% step 3 mean 0.1776 error 0.1206
% step 5 mean 0.1287 error 0.1005

\nocite{*}
\bibliography{aipsamp}% Produces the bibliography via BibTeX.

\end{document}